\documentclass[aps,prc,twocolumn,showpacs,groupedaddress,superscriptaddress
]{revtex4-1}
\usepackage{graphicx}% Include figure files
\usepackage{amsmath} %for text inside the equation, 
\usepackage{amssymb}
\usepackage{amsfonts}
\usepackage{mathrsfs}
\usepackage{dcolumn}% Align table columns on decimal point
\usepackage{bm}% bold math
\usepackage{rotating} %Sideways tables
\usepackage{MnSymbol}
\usepackage{color}
\usepackage[dvipsnames]{xcolor}
\usepackage[active]{srcltx}
\usepackage{multirow}
\usepackage{caption}
\usepackage{subcaption} %For subfigures
\usepackage{comment} % For multiple comment lines
\usepackage[normalem]{ulem}

\newcommand{\si}[1]{\,\rm{#1}}
\newcommand{\sn}[1]{\times10^{#1}}
\newcommand{\nn}[2]{^{#1}\rm{#2}}

\begin{document}
\title{Alpha-decay from $^{44}$Ti: A study of the microscopic clusterization}

%-----------------------
\author{A. C. Dassie}
\affiliation{Instituto de F\'isica Rosario (CONICET-UNR), Ocampo y Esmeralda, Rosario 2000. Argentina.}
\affiliation{Facultad de Ciencias Exactas, Ingenier\'ia y Agrimensura (UNR), Av. Pellegrini 250, Rosario 2000. Argentina.}
%-----------------------
\author{R. M. Id Betan}
\affiliation{Instituto de F\'isica Rosario (CONICET-UNR), Ocampo y Esmeralda, Rosario 2000. Argentina.}
\affiliation{Facultad de Ciencias Exactas, Ingenier\'ia y Agrimensura (UNR), Av. Pellegrini 250, Rosario 2000. Argentina.}

%____________
\date{\today}

%_______________
\begin{abstract}  
Microscopic determination of alpha-decay half-lives requires structure and reaction calculations. The structure part is given by the microscopic distribution of the constituent nucleons, while the relative motion of the product's decay provides the reaction part. This paper studies the clusterization of the $0^+$ excited states of $^{44}$Ti arising from the nucleonic degrees of freedom. The continuum spectra of proton and neutron are incorporated through the Gamow Shell Model formalism. The alpha-like wave function is calculated in the weak-coupling approximation. Gaussian effective interaction in each pair of nucleons is included. The $0^+$ ground and excited states are compared with experiment and shell model calculations. The wave function amplitudes are obtained and discriminated by their resonant and non-resonant contributions. The influence of a four-body truncated basis is analyzed. Inclusion of the continuum spectra produces a gain in the excited states of a few hundred keV of energy. One candidate for alpha decay was identified near the experimental unresolved state $(0,2)^+$ of 6.8 MeV excitation energy, with a lower limit for the half-life of $\approx 0.8$ ns.
\end{abstract}
 
%_____________________________________________
% 21.10.-k 	Properties of nuclei; nuclear energy levels
% 21.10.Dr 	Binding energies and masses
% 21.10.Ma 	Level density
% 21.10.Pc 	Single-particle levels and strength functions
% 21.30.Fe 	Forces in hadronic systems and effective interactions
% 21.60.Cs 	Shell model
% 27. 	Properties of specific nuclei listed by mass ranges
% 27.40.+z	39 ≤ A ≤ 58
% 27.80.+w 	190 ≤ A ≤ 219

%_________
\maketitle

%=============================================
\section{Introduction}\label{sec.introduction}
A unified treatment of the alpha-decay from the shell model framework is possible \cite{1957Mang,1979Tonozuka}. On one hand formation process involves the structure calculation of a many-body overlap between the mother nucleus and the product's decay. On the other hand, the penetration process needs to appeal to a resonant state. In general terms, processes relate to different aspects of nuclear physics; the former belongs to nuclear structure and the latter to nuclear reaction. In Ref. \cite{2012IdBetan}, a unified framework for alpha-decay calculations was presented in the pole approximation of the Gamow shell model. In the present work the non-resonant continuum is incorporated into the single-particle representation. An effective Gaussian interaction replaces the separable force. The coulomb interaction between the valence proton is also considered. The missing proton-neutron interaction is now taken into account. While the two-body mean-field parameters are constrained using low-lying excited states from neutron-neutron, proton-proton, and neutron-proton experimental data. Finally, the alpha-like wave function is constructed using the weak-coupling scheme \cite{1965Glendenning,1980Lawson}. 

We consider $\nn{44}{Ti}$ as our case study, in which evidence of $\alpha$ structure has been observed through multiple reactions, also at excitation energies above the alpha threshold \cite{frekersZPhysikA1976,frekersNuclearPhysicsA1983,yamayaPhys.Rev.C1990}.
 It is also a nucleus of significant astrophysical interest because of its implications in core-collapse supernovae 
\cite{theApJ1998,nassarPhys.Rev.Lett.2006,vockenhuberPhys.Rev.C2007,larsenPhys.Rev.C2012}. At the same time, evidence of $\alpha$ structure has been observed through multiple reactions, also at excitation energies above the alpha threshold \cite{frekersZPhysikA1976,frekersNuclearPhysicsA1983,yamayaPhys.Rev.C1990}. This paper presents the treatment of the full continuum correlation on alpha clusterization from the nucleon degree of freedom. The following study will present the microscopic calculation of the alpha-decay half live using the microscopic spectroscopic factor. 

Section \ref{sec.formalism} presents the two-step process \cite{1965Glendenning} to define the four-body basis. Section \ref{sec.representation} defines the single particle representation, while Sec. \ref{sec.vtwo} describes the isospin-dependent two-body interactions. Section \ref{sec.0state} studies the influence of the continuum over the four-body $0^+$ spectrum and the wave function, while Sec. \ref{sec.trunc} shows the effects on the collectivity due to truncation. Sec. \ref{sec.t} calculates the half-lives in the two-body approximation. Finally Sec. \ref{sec.conclusions} summarizes the results and outlines the next step related to the alpha decay in $^{44}$Ti.

%=======================================
\section{Four-body wave function}\label{sec.formalism}
This section presents the two-step process \cite{1965Glendenning,1974True,1980Lawson} formalism, which provides a systematic way of truncating the four-body basis. Correlated two-like nucleon bases are formed first by diagonalizing the two-like nucleon parts of the Hamiltonian. Each one of these two-body eigenfunctions is expanded in the uncorrelated two-like nucleon basis formed by the eigenfunctions of the core-nucleon parts of the five-body Hamiltonian,
\begin{equation*}
  \mathcal{H}=H_{n}+H_{p}+V_{np} \, .
\end{equation*}

The alpha-like wave function is expanded in a basis built from the neutron-neutron $\Psi_{J_n^\pi M_n}$ and proton-proton $\Psi_{J_p^\pi M_p}$ eigenfunctions of the Hamiltonian of each pair of the two-like nucleon, respectively,
\begin{align*}
    H_n \Psi_{J_n^\pi M_n} &= E_{J_n^\pi} \Psi_{J_n^\pi M_n} \, , \\
    H_p \Psi_{J_p^\pi M_p} &= E_{J_p^\pi} \Psi_{J_p^\pi M_p} \, .
\end{align*}

The two-like nucleon Hamiltonians are:
\begin{align*}
   H_{n} &= h_n(\bar{r}_1) + h_n(\bar{r}_2) + V(\bar{r}_1,\bar{r}_2)\, , \\
   H_{p} &= h_p(\bar{r}_3) + h_p(\bar{r}_4) + V(\bar{r}_3,\bar{r}_4) 
   + \frac{e^2}{|\bar{r}_3-\bar{r}_4|} \, ,
\end{align*}
while $V_{np}=V(\bar{r}_1,\bar{r}_3)+V(\bar{r}_1,\bar{r}_4)+V(\bar{r}_2,\bar{r}_4)+V(\bar{r}_2,\bar{r}_4)$ is the remaining proton-neutron interaction.

An isospin-dependent effective interaction in all nucleon-nucleon channels is used,
\begin{equation}
  V(r) = \sum_\tau V^J_\tau(r) P_\tau 
\end{equation}
with $r=|\bar{r}_i-\bar{r}_j|$, $P_\tau$ the projector operator over one of the spin-isospin channel $\tau= \{ se,\, to,\, se,\, so \}$, and
\begin{equation}\label{eq.2}
   V^J_{\tau}(r) = V^J_{\tau} e^{-\frac{r^2}{\beta^2_\tau}} \, .
\end{equation}

The neutron and proton single-particle mean-field Hamiltonians contain the central Woods-Saxon, the spin-orbit interactions, and the Coulomb one for the protons. Bound, resonances and complex energy scattering eigenstates are calculated from the single-particle Hamiltonian,
\begin{equation}
  h(\bar{r})  \psi_{a,m_a}(\bar{r}) 
   = \varepsilon_a \psi_{a,m_a}(\bar{r})
\end{equation}
with $a=\{ n_a, l_a, j_a\}$.

The bound and continuum single-particle states are used to generate the proton-proton and neutron-neutron bases \cite{1968Berggren,1996Liotta,2002IdBetan,2002Michel}
\begin{equation} \label{eq.psitwo}
   \Psi_{J^\pi M} = \sum_{a \le b} X^{J^\pi}_{ab} \Psi^{J^\pi M}_{ab} \, 
\end{equation}
with the amplitude $X^{J^\pi}_{ab}$ normalized within the Berggren metric, i.e. $\sum_{a \le b} (X^{J^\pi}_{ab})^2=1$.

Then, the correlated alpha-like wave function reads
\begin{equation} \label{eq.four} 
   | J M \rangle = \sum_{J_{n} J_{p}} 
        Z^{J}_{np} |J_n J_p,J \rangle \, ,
\end{equation}
where $|J_n J_p,J \rangle=[\Psi_{J_n^\pi} \Psi_{J_p^\pi}]_{J^\pi M}$.

In the adopted weak-coupling interaction, the secular equation contains only diagonal elements in the nucleon-like quadrants
\begin{align*} 
   \sum_{J'_{n} J'_{p}}
   &\left[ (E_{J_{n}} + E_{J_{p}}) 
   \delta_{J'_{n} J_{n}}
   \delta_{J'_{p} J_{p}} 
   \right. \nonumber \\
    &\left.        
    + \langle J_{n} J_{p}, J^\pi | V_{np} | J'_{n} J'_{p}, J^\pi \rangle   
   \right] Z^{J}_{J'_{n} J'_{p}} 
    = E_{J} Z^{J}_{J_{n} J_{p}}  \, ,
\end{align*}
with $E_{J}$ the alpha-like eigenenergy $\mathcal{H}|JM\rangle=E_J|JM\rangle$.

%============================================
\section{Application} \label{sec.application}

%--------------------------
\subsection{Representations}\label{sec.representation}
To separately assess the influence of the resonant and non-resonant continuum in the alpha-like wave functions, we define three single-particle representations called: (i) \textit{bound basis} (BB), (ii) \textit{pole basis} (PB), and (iii) \textit{complete basis} (CB). The election of each one of these single-particle bases will impact the neutron-neutron and proton-proton bases and, consequently, in the four-body basis. In this way, we can quantify the contribution of the continuum on the many-body calculation.

The \textit{bound basis} contains only neutron and proton states bound to the core. Then, the alpha-like wave functions will not include correlations with the continuum part of the energy spectra, neither the neutron nor the proton ones. The \textit{pole basis} contains, besides the bound states, the resonant states of the core-nucleon systems. Since the scattering contours are absent, the completeness of the Berggren representation is broken. Consequently, small imaginary parts may appear in wave function amplitudes and energies, also at the two-particle stage. Then, the four-body amplitudes will contain configurations that partially include the continuum. 
The third basis, the \textit{complete basis}, also includes the non-resonant continuum, particularly the complex contours companion of the resonances included in the pole basis. This basis restores the completeness of the complex energy representation, and former imaginary components of physically real magnitudes, became zero (to some numerical resolution). 

The core-nucleon mean-field parameters are optimized to the energies of the nuclei $^{41}$Ca and $^{41}$Sc from Ref. \cite{2007Schwierz}. 
The strengths and diffuseness are optimized by using $\chi^2$ minimization, with the reduced radius fixed by the experimental nucleon root-mean-square radius $r_n(p)=3.375(3.385)$ fm \cite{2018Zenihiro}. The same diffuseness and reduced radius are used for the Woods-Saxon and the spin-orbit mean fields. A Coulomb potential of uniform charge distribution is added to the proton interaction with the same radius as for the Woods-Saxon. The parameters are shown in Table \ref{table.vsp}. 
%-----------------
\begin{table}[h!t]
\centering
\caption{\label{table.vsp} Neutron and proton parameters (errors in parenthesis) for the Woods-Saxon and spin-orbit mean-fields \cite{1982Vertse}.} 
\begin{tabular}{c|cccc}
    \hline
    nucleon & $V_0$(MeV) & $V_{so}$(MeV) &  $a$(fm) & $r_0$(fm) \\
    \hline
    neutron & $52.052(1)$ & $16.915(4)$ & $0.811(0.2)$ & $1.274$ \\
    proton  & $51.427(1)$ & $16.191(5)$ & $0.791(0.2)$ & $1.278$ \\
    \hline
\end{tabular}
\end{table}

The neutron and proton single-particle states are calculated using the code GAMOW \cite{1982Vertse}. Table \ref{table.spbase} shows them for the first two major shells above the $^{40}$Ca.
%-----------------
\begin{table}[h!t]
\centering
\caption{\label{table.spbase} Neutron ($\nu)$ and proton ($\pi$) energies (MeV) in $^{40}$Ca.}
\begin{tabular}{c|cc|cc}
  \hline 
          & \multicolumn{2}{c|}{$\varepsilon_\nu$} 
          & \multicolumn{2}{c}{$\varepsilon_\pi$} \\                    
    state & GAMOW  &  Ref. \cite{2007Schwierz} 	
          & GAMOW  &  Ref. \cite{2007Schwierz} \\ 
    \hline
    $0f_{7/2}$  & -8.309 			&-8.36& -1.109 &-1.09\\
    $1p_{3/2}$  & -6.017			&-5.84& (0.760,-0.007$\sn{-3}$)& 0.69\\
    $1p_{1/2}$  & -3.995			&-4.20& (2.291,-0.049) &2.38\\
    $0f_{5/2}$  & -1.626          &-1.56& (4.982,-0.079) &4.96\\			
    $0g_{9/2}$  & (1.658,-0.004)  && (7.958,-0.222) &\\
    $0g_{7/2}$  & (8.321,-1.541)  && (14.430,-2.787)& \\
    $2d_{5/2}$  & (0.895,-0.188)  && (6.065,-1.734) &\\
    $2d_{3/2}$  & (1.954,-1.335)  && (6.951,-3.847) &\\
    $0h_{11/2}$ & (10.568,-1.129) && (16.554,-2.074) & \\
    $0h_{9/2}$  & (17.876,-7.300) && (24.164,-9.193)&  \\ 
    \hline
\end{tabular}
\end{table}

The neutron and proton bound bases (BB) include the real energy states of Table \ref{table.spbase}. The pole basis (PB) also incorporates the resonances with $|\Im(\varepsilon)| < 0.25$ MeV, i.e., $\{ 2d_{5/2},0g_{9/2} \}$ neutron states, and $\{ 1p_{3/2},1p_{1/2},0f_{5/2},0g_{9/2} \}$ proton states. Finally, a discretized number of Gauss-Legendre real or complex energy scattering states are added to the PB to generate the complete bases (CB). Triangular-shaped complex contours \cite{2002Michel} are defined to enclose the neutron $\{ 2d_{5/2},0g_{9/2} \}$ and proton $\{ 1p_{3/2},1p_{1/2},0f_{5/2},0g_{9/2} \}$ resonances, respectively. The contours are separated enough from the poles such that they do not interfere with each other \cite{2003IdBetan}. Neutron $s$ and $p$, and proton $s$ real energy scattering states are also included in the CB. The energy cutoff for all partial waves is taken as 12 MeV. Table \ref{table.4basis} summarizes each one of the neutron and proton bases considered. 
%----------------
\begin{table}[h!t]
\centering
\caption{\label{table.4basis} (color online) Bound (BB), pole (PB), and complete (CB) single particle bases, with $c$ preceding the scattering partial waves. Complex contours appear in blue.} 
\begin{tabular}{c|c|c}
    \hline
    Basis  & Neutron states 
    & Proton states \\
    \hline
    BB &  $0f_{7/2}$, $1p_{3/2}$, $1p_{1/2}$, $0f_{5/2}$ & $0f_{7/2}$ \\
    \hline
    \multirow{2}{*}{PB} 
    & \multirow{2}{*}{BB + {\color{blue}$2d_{5/2}$},{\color{blue}$0g_{9/2}$}} & BB + {\color{blue}$1p_{3/2}$}, {\color{blue}$1p_{1/2}$} \\
    &  & {\color{blue}$0f_{5/2}$}, {\color{blue}$0g_{9/2}$} \\
    \hline
    \multirow{2}{*}{CB} 
    & PB + {\color{blue}$cd_{5/2}$}, {\color{blue}$cg_{9/2}$}, $cs_{1/2}$ & PB + {\color{blue}$cp_{3/2}$}, {\color{blue}$cp_{1/2}$} \\
    & $cp_{1/2}$, $cp_{3/2}$ & {\color{blue}$cf_{5/2}$}, {\color{blue}$cg_{9/2}$}, $cs_{1/2}$ \\
    \hline
\end{tabular}
\end{table}

%---------------------------------
\subsection{Two-body interactions}\label{sec.vtwo}
The two-body residual interaction \eqref{eq.2} is separately optimized for the $T=1$ and $T=0$ channels using the Levenberg-Marquardt $\chi^2$ algorithm \cite{nr}. The range of the interaction is taken as $\beta=1.6\si{fm}$ \cite{Newby1959} for all channels. A first optimization using the even low-lying states of the three nuclei $^{42}$Ca, $^{42}$Sc, and $^{42}$Ti gives a residue of about 800 keV. This figure is much reduced (around 30 keV) considering only the even states of Table \ref{table.exp}. Although the ground state of $^{42}$Sc departs 1.4 MeV from the experimental one, this mean field is more convenient to generate the correlated bases. The triplet-odd strength mildly influences the spectra, so it was taken to be zero \cite{True1958,Newby1959,Kim1963,1965Glendenning}. Finally, the $T=0$ strengths are optimized using the odd $^{42}$Sc states listed in Table \ref{table.exp}.
%----------------
\begin{table}[h!t]
\centering
\caption{\label{table.exp} Experimental \cite{nndc} and calculated low-lying energies (MeV) in the $^{40}$Ca plus two nucleons. BB, PB, and CB refer to the single-particle model space from which the two-body wave functions are expanded.}
\begin{tabular}{cc|ccc}
    $J^\pi$ & Exp. & BB & PB & CB \\
    \hline \hline 
    & \multicolumn{4}{c}{$^{42}$Ca} \\
    \hline
    $0^+$   & $-19.843$ & $-20.093$ & $-19.923+i0.016$ & $-19.925$ \\
    $2^+$   & $-18.319$ & $-18.428$ & $-18.293+i0.010$ & $-18.291$ \\
    \hline 	\hline
    & \multicolumn{4}{c}{$^{42}$Ti}\\
    \hline
    $0^+$   & $-4.836$ & $-4.502$ & $-4.745+i0.098$ & $-4.741$ \\
    $2^+$   & $-3.282$ & $-3.034$ & $-3.311+i0.040$ & $-3.316$ \\
    \hline 	\hline
    & \multicolumn{4}{c}{$^{42}$Sc}  \\
    \hline
    $1^+$   & $-9.799$  & $-9.799$  & $-9.803+i0.065$  & $-9.801$ \\
    $7^+$   & $-9.795$  & $-9.790$  & $-9.794$ & $-9.794$ \\
    \hline
\end{tabular}
\end{table}

The optimization is carried out for each one of the single particle representations BB, PB, and CB. Table \ref{table.veff} shows the optimized strengths, while in the previous Table \ref{table.exp}, we compare the calculated energies with the experimental ones. Due to the missing non-resonant continuum in the pole basis, the calculated energies using this representation show an spurious imaginary component \cite{1993Beggren,1989Curutchet}. Using 30 neutron and 32 proton non-resonant continuum states in the CB, the real character of the two-body energies is restored.
%------------------
\begin{table}[h!t]
\centering
\caption{\label{table.veff} Optimized two-nucleon strengths (MeV) used for calculating the two-body correlated bases with $\beta=1.6\si{fm}$.}
\begin{tabular}{ c | c | c c c}
    \hline
    $V_\tau$ & $J$  & BB & PB & CB \\
    \hline
    \multirow{2}{*}{$se$} & $=0$ & $-58.823$ & $-52.623$ & $-51.023$ \\
                          & $>0$ & $-73.116$ & $-66.497$ & $-63.816$ \\
    \hline
    $so$                  &      & $-5.479$ & $-5.125$ & $-5.059$ \\
    $te$                  &      & $-9.595$ & $-9.672$ & $-9.659$ \\
    \hline
\end{tabular}
\end{table}

\begin{figure}[h!t]
    \centering
    \includegraphics[width=0.9\columnwidth]{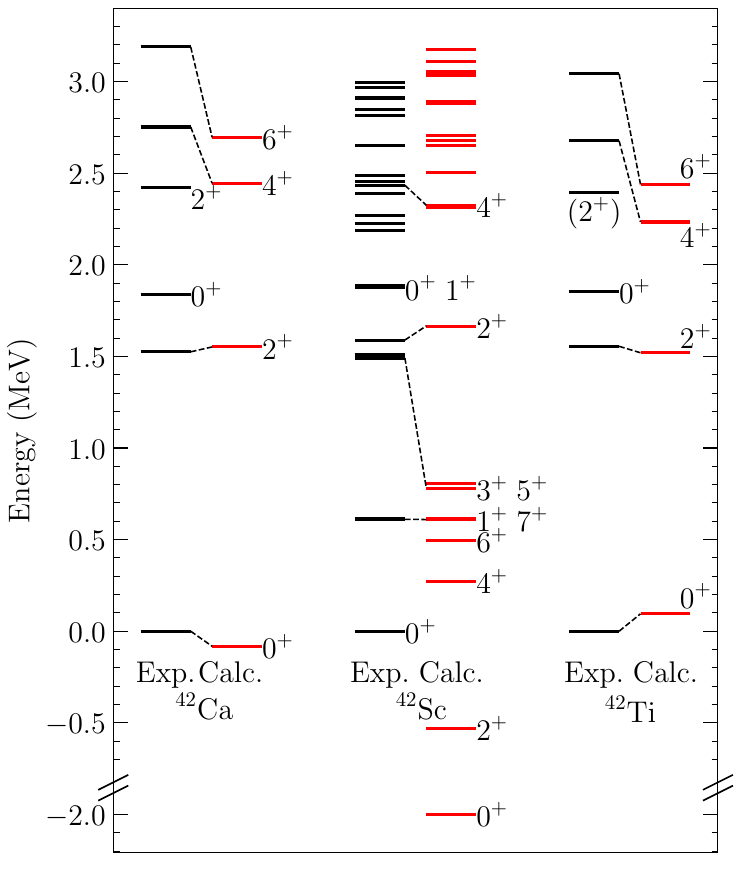}
    \caption{Experimental and calculated two body energies for the Complete basis.}
    \label{fig:tbe}
\end{figure}

Figure \ref{fig:tbe} compares the low-lying part of the experimental spectra of $\nn{42}{Ca}$, $\nn{42}{Sc}$, and $\nn{42}{Ti}$ with the calculated ones using the complete basis. To describe more precisely the four-body threshold, we consider different strengths for $J=0$. The calculated states with $J>2$ are over-bound, indicating that smaller strengths are needed. To keep the number of free parameters as small as possible, we stick with the parameter of Table \ref{table.veff} to generate the neutron-neutron and proton-proton bases.
%-------------------

%-------------------------------------
\subsection{$0^+$ states in $^{44}$Ti }\label{sec.0state}
In this section, the $0^+$ states of the nucleus $^{44}$Ti are calculated, and the amount of clusterization is assessed from the collective character of the wave function. Comparison with a truncated two-body basis is performed. The contents of the continuum on the wave functions are analyzed. 

Each single particle basis BB, PB, and CB generates a two-nucleon basis, which in turn lets to three four-body bases. Likewise the single particle basis, we label the result for the four body calculation using the same labelling, i.e., BB, PB, and CB. Since each one of the nucleon-nucleon correlations are taken into account, the content of continuum four-body correlations between neutron-neutron, proton-proton, and proton-neutron increment sequentially from the BB to the CB bases. In particular, for the complete basis we keep configurations which contains at most three nucleons in the non-resonant continuum \cite{2017Jaganathen}.
 
The ground-state energy of the $^{42}{\rm Ca}$ plus that of $^{42}\rm{Ti}$ gives the uncorrelated four-body ground-state energy of $-24.679$ MeV, while the experimental ground state energy of $^{44}$Ti is $-33.423$ MeV. Then, the proton-neutron interaction has to provide $8.744$ MeV of correlation energy. Using the proton-neutron strength of Table \ref{table.veff} in the four-body Hamiltonian, the gain is around 2 MeV. To get the experimental energy we increase the strengths of $V_{so}$ and $V_{te}$ by the factors $\chi_{\rm{BB}}=7.7028$, $\chi_{\rm{PB}}=7.8191$, and $\chi_{\rm{CB}}=7.9169$, for each one of the basis, respectively. We may interpret that the parameter $\chi$ measures the increase of the proton-neutron $T=0$ part of the correlations in the four-body medium. This parameter also indicates the viability of the two-proton two-neutron approach without including the proton-neutron interaction to describe alpha-like states in medium size nuclei.

Table \ref{table.wfb} shows the low-lying $0^+$ states of $^{44}$Ti calculated using each BB, PB, and CB bases. The two-body model spaces include all bound states, i.e., two-proton states with energy up to $S^{\mathrm{calc}}_{\rm{2p}}(\,\nn{42}{Ti})$, and two-neutron states up to $S^{\mathrm{calc}}_{\rm{2n}}(\,\nn{42}{Ca})$. Table \ref{table.wfb} shows the amplitude of the main configurations. All the exited states are sitting above the alpha threshold $\alpha+^{40}$Ca threshold, $-28.301\si{MeV}$ ($Q_\alpha=-5.127$ MeV). The calculation from the PB has an imaginary component that vanishes in the CB. Except for the third excited state, when changing from the PB to the CB, the inclusion of the continuum produce a gain in the correlations. The average over-binding energy from the pole to the complete continuum is about 200 keV, except for the fourth and fifth states that are less bound.
%-----------------
\begin{table}[h!t]
\renewcommand*{\arraystretch}{1.4}
\centering
\caption{\label{table.wfb} Low-lying $0^+$ energies and wave functions of $^{44}$Ti calculated using the BB, PB, and CB model spaces. $\mathcal{Z}^2_{J^\pi}$ are the four-body amplitudes, 
$\mathcal{Z}^2_{J^\pi} =\sum_{i} (Z^{0^+}_{i})^2$, with $Z^{0^+}_{i}=Z^{0^+}_{np}$ of Eq. \eqref{eq.four}, and $i$ labeling all configurations with $[J_n,J_p]_{J^\pi}$.
} 
\begin{tabular}{c|ccc}
    \hline
	$\mathcal{Z}^2_{J}$ & BB & PB & CB  \\ 
	\hline
	$E_{0^+_1}$ & $-33.423$	& $(-33.423,0.197)$ & $-33.423$  \\[0.3mm]
	\hline
	$\mathcal{Z}^2_{0^+}$	& $0.616$ & $(0.616,0.000)$ & $0.609$ \\
	$\mathcal{Z}^2_{2^+}$	& $0.320$ & $(0.325,0.000)$ & $0.330$ \\
	$\mathcal{Z}^2_{4^+}$	& $0.057$ & $(0.053,0.001)$ & $0.054$ \\
	$\mathcal{Z}^2_{6^+}$	& $0.007$ & $(0.005,0.000)$ & $0.005$ \\
	\hline
    $\sum_J \mathcal{Z}^2_J$ & $1$ & $0.999$ & $0.998$ \\
	\hline
	\hline
	$E_{0^+_2}$ & $-27.338$ & $(-27.888,0.155)$ & $-28.086$  \\[0.3mm]
	\hline
	$\mathcal{Z}^2_{0^+}$	& $0.164$ & $(0.079,0.002)$ & $0.087$ \\
	$\mathcal{Z}^2_{2^+}$	& $0.496$ & $(0.655,0.016)$ & $0.708$ \\
	$\mathcal{Z}^2_{4^+}$	& $0.022$ & $(0.079,0.002)$ & $0.086$ \\
	$\mathcal{Z}^2_{6^+}$	& $0.318$ & $(0.180,0.030)$ & $0.112$ \\
	\hline
    $\sum_J \mathcal{Z}^2_J$ & $1$ & $0.993$ & $0.993$ \\
	\hline
	\hline
	$E_{0^+_3}$ & $-25.755$ & $(-26.640,0.136)$	& $-26.782$ \\[0.3mm]
	\hline
	$\mathcal{Z}^2_{0^+}$	& $0.190$ & $(0.207,-0.009)$ & $0.222$ \\
	$\mathcal{Z}^2_{2^+}$	& $0.130$ & $(0.470,-0.002)$ & $0.424$ \\
	$\mathcal{Z}^2_{4^+}$	& $0.128$ & $(0.047,0.004)$ & $0.032$ \\
	$\mathcal{Z}^2_{6^+}$	& $0.552$ & $(0.249,0.000)$ & $0.296$ \\
	\hline
    $\sum_J \mathcal{Z}^2_J$ & $1$ & $0.973$ & $0.974$ \\
	\hline
	\hline
	$E_{0^+_4}$ & $-25.569$	& $(-25.574,0.073)$	& $-25.566$ \\[0.3mm]
	\hline
	$\mathcal{Z}^2_{0^+}$	& $0.016$ & $(0.036,-0.002)$ & $0.022$ \\
	$\mathcal{Z}^2_{2^+}$	& $0.078$ & $(0.211,-0.007)$ & $0.175$ \\
	$\mathcal{Z^2}_{4^+}$	& $0.810$ & $(0.542,-0.002)$ & $0.662$ \\
	$\mathcal{Z}^2_{6^+}$	& $0.096$ & $(0.202,-0.034)$ & $0.135$ \\
	\hline
    $\sum_J \mathcal{Z}^2_J$ & $1$ & $0.991$ & $0.994$ \\
	\hline
	\hline
	$E_{0^+_5}$ & $-24.923$ & $(-25.346,0.079)$	& $-25.232$ \\[0.3mm]
	\hline
	$\mathcal{Z}^2_{0^+}$	& $0.334$ & $(0.209,0.019)$ & $0.240$ \\
	$\mathcal{Z}^2_{2^+}$	& $0.524$ & $(0.107,0.016)$ & $0.157$ \\
	$\mathcal{Z}^2_{4^+}$	& $0.127$ & $(0.387,0.000)$ & $0.245$ \\
	$\mathcal{Z}^2_{6^+}$	& $0.015$ & $(0.284,0.023)$ & $0.352$ \\
	\hline
    $\sum_J \mathcal{Z}^2_J$ & $1$ & $0.987$ & $0.994$ \\
	\hline
	\hline
	$E_{0^+_6}$ & $-22.246$	& $(-23.799,0.139)$ & $-24.022$ \\[0.3mm]
	\hline
	$\mathcal{Z}^2_{0^+}$	& $0.417$ & $(0.368,0.000)$ & $0.372$ \\
	$\mathcal{Z}^2_{2^+}$	& $0.262$ & $(0.458,0.012)$ & $0.437$ \\
	$\mathcal{Z}^2_{4^+}$	& $0.211$ & $(0.135,0.001)$ & $0.140$ \\
	$\mathcal{Z}^2_{6^+}$	& $0.110$ & $(0.011,0.000)$ & $0.018$ \\
	\hline
    $\sum_J \mathcal{Z}^2_J$ & $1$ & $0.972$ & $0.967$ \\
	\hline
\end{tabular}
\end{table}

\begin{figure}[h!t]
    \centering
    \includegraphics[width=0.49\textwidth]{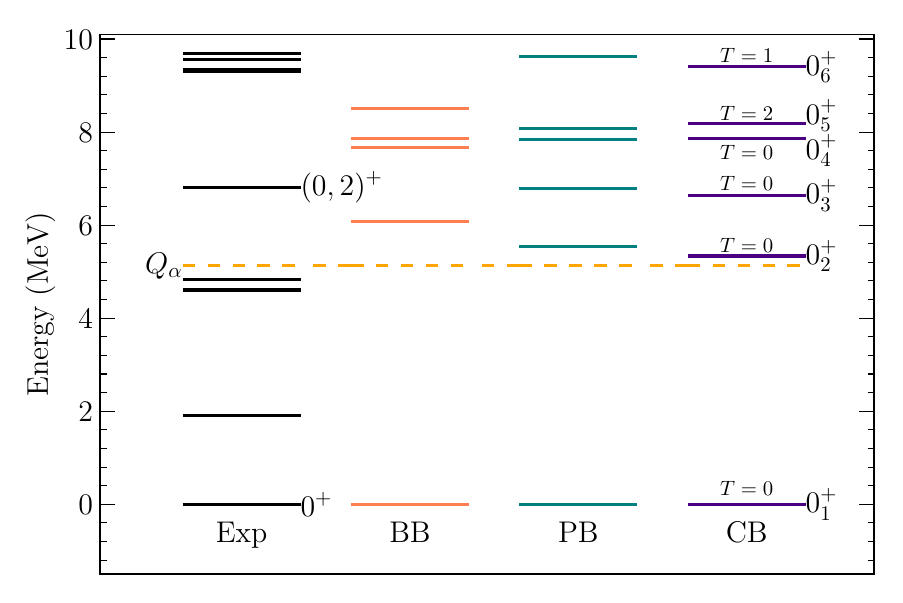}
  \caption{Low-lying $0^+$ states of $^{44}$Ti. The results for the three single-particle model spaces are shown. The experimental values are from Ref. \cite{nndc}.}
  \label{fig.exp0}
\end{figure}

Figure \ref{fig.exp0} shows the experimental and calculated $0^+$ levels of $^{44}$Ti labeled with their corresponding isospin. Our model calculation does not find any of the first three excited states below the alpha threshold due to the truncation of the model space \cite{1969Shah}. In particular, the states at 1.905 MeV and 4.9 MeV were not found neither in Ref. \cite{Dixon1978}. The first calculated exited $0^+_2$ state at the energy $5.337$ MeV agrees with the one estimated in Ref. \cite{1969Shah}. In this Ref., only states with the same configuration were kept, $a=b$ in Eq. \eqref{eq.psitwo}. The isospin assignments and energies of the states above the alpha threshold agree with the ones calculated using the complete model space of Ref. \cite{Dixon1978}. Still, we get two fewer states near the $T=2$ ones. The adequacy of the truncation was checked by incrementally adding two-body states to the basis. It was found that the energy position converged to the values shown in Fig. \ref{fig.exp0}.  

The $0^+_2$ is only $210$ keV above the alpha threshold, at the excitation energy of $5.337$ MeV. None of the states listed in \cite{nndc} around 5 MeV is a $0^+$ state. Reference \cite{Dixon1977} lists a state at the energy $5.304$ or $5.315$ (values came from different reactions). This state may correspond to the calculated $0^+_2$ state. The $0^+_3$ is nearby of the unconfirmed $(0,2)^+$ state, while the $0^+_6$, at the excitation energy $9.401$ MeV, is in between the six experimental $0^+$ states with energy in the range $9.14-9.78$ MeV. 
%-------------------

Besides the energies, table \ref{table.wfb} also shows the contribution to the norm of the alpha-like wave function from configuration $[J_n,J_p]_{0^+}$. The figures in the Table include the sum of different energy states with the same angular momentum. The four-body basis states in each basis are $20$, $78$, and $1437$ for the BB, PB, and CB, respectively. The ground-state is built mainly from the $0^+$ and $2^+$ states of the correlated $^{42}$Ca and $^{42}$Ti. The energies and amplitudes may significantly be affected by the inclusion of the resonances, while when adding the non-resonant continuum, the real part of the amplitudes are mildly affected, but the imaginary component is cancelled out. The state $0^+_5$ is the only one with a non-negligible contribution from each configuration $[J_n,J_p]_{J^\pi}$. Next, the state $0^+_3$ has important contributions but the two-body $4^+$ configuration. These two states may be candidates for alpha decay from $^{44}$Ti. The collectively is only one aspect in the alpha decay calculation \cite{2012IdBetan}; the other, the single particle width, will be analyzed in the next section.

The Berggren representation allows discriminating the content of the single-particle continuum in the many-body wave function \cite{2004IdBetan}. Let us define the bound-bound (B-B) contribution to the norm as the sum of four-body configurations when both neutrons and protons are in bound states. The bound-resonant (B-R) probability sums up all single-particle configurations for which at least one of the nucleon is in a resonant state and the others in bound configurations. Likewise, the pole-continuum (P-C) probability contains all configurations for which at least one of the nucleons is sitting in the non-resonant continuum and the others in any of the poles (bound or resonant states). Finally, the continuum-continuum probability sums up the remainder amplitudes. Table \ref{table.wfb2} shows the amplitudes of Table \ref{table.wfb} (last column) separated by the different single-particle bases, to assess the influence of the continua spectra (proton and neutron) in the many-body calculation. One can observe that the continuum makes more than 10\% of the configurations, with the major contribution coming from the mixing between bound and resonant configurations.
%------------------
\begin{table*}[h!t]
	\centering
	\renewcommand*{\arraystretch}{1.4}
	\caption{\label{table.wfb2}Four-body wave function amplitudes for the $0^+$ states discriminated by single-particle configurations, as explained in the text. The Continuum-Continuum contributions are of the order of the eV.}
	%\resizebox{\textwidth}{!}{
	\begin{tabular}{c|ccc|ccc|ccc|ccc|ccc}
		\hline
        $E_{0^+}$ (MeV) 					& \multicolumn{3}{c|}{$-33.423$} & \multicolumn{3}{c|}{$-28.086$}
        & \multicolumn{3}{c|}{$-26.782$} & \multicolumn{3}{c|}{$-25.566$} & \multicolumn{3}{c}{$-25.232$}\\
        \hline
        $\mathcal{Z}^2_{J^\pi}$ 	
        & B-B & B-R & P-C
        & B-B & B-R & P-C
        & B-B & B-R & P-C
        & B-B & B-R & P-C
        & B-B & B-R & P-C\\
        \hline
        $0^+$ 
        & $0.550$ & $0.041$ & $0.002$ 
        & $0.077$ & $0.005$ & $0.002$ 
        & $0.200$ & $0.014$ & $0.002$ 
        & $0.019$ & $0.001$ & $0$ 
        & $0.216$ & $0.016$ & $0.001$ \\
        
        $2^+$ 
        & $0.264$ & $0.058$ & $0.003$ 
        & $0.390$ & $0.300$ & $0.007$ 
        & $0.285$ & $0.129$ & $0.004$ 
        & $0.122$ & $0.049$ & $0.001$ 
        & $0.129$ & $0.024$ & $0.001$ \\
        
        $4^+$ 
        & $0.049$ & $0$     & $0$     
        & $0.053$ & $0.031$ & $0.001$ 
        & $0.024$ & $0.007$ & $0$     
        & $0.603$ & $0.050$ & $0.003$ 
        & $0.222$ & $0.020$ & $0.001$ \\ 
        
        $6^+$ 
        & $0$     & $0$     & $0$     
        & $0.114$ & $0.003$ & $0$     
        & $0.286$ & $0.008$ & $0$     
        & $0.130$ & $0.004$ & $0$     
        & $0.340$ & $0.009$ & $0$ \\
        
        \hline
        $\sum_J \mathcal{Z}^2_{J}$
        & $0.868$ & $0.105$ & $0.005$ 
        & $0.634$ & $0.341$ & $0.010$ 
        & $0.795$ & $0.183$ & $0.006$ 
        & $0.874$ & $0.111$ & $0.004$ 
        & $0.908$ & $0.075$ & $0.003$ \\
        \hline
    \end{tabular}%}
\end{table*}

\subsection{Four-body truncated basis}\label{sec.trunc}
\begin{figure}[h!t]
    \centering
    \includegraphics[width=0.9\columnwidth]{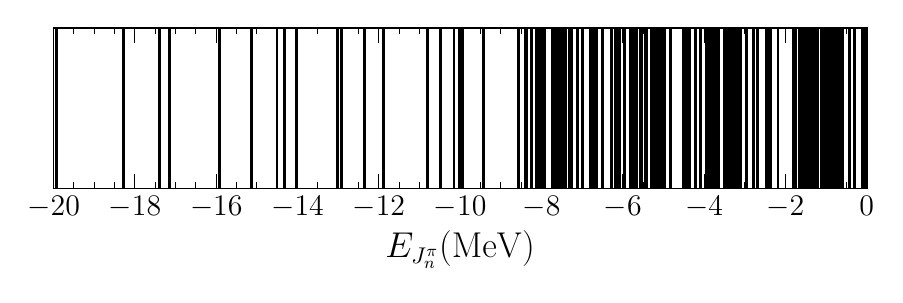}
    \caption{Two-neutron correlated bound states. The quantum number $E_{J^\pi}$ of each line is not stated for simplicity.}
    \label{fig:tne}
\end{figure}

The analysis of the four-body collectivity requires the calculation of the formation amplitude \cite{2012IdBetan}, which implies a large number of many-dimensional integrals. However, it is possible to truncate the two-body bases to reduce this calculation without significantly losing the collective character of the state. Figure \ref{fig:tne} shows the correlated two-neutron bases corresponding to the complete single-particle basis. The massive growth of the two-body basis states above $-9$ MeV suggests defining a two-neutron basis with states up to $10$ MeV of excitation energy.
With this basis, the number of four-body basis states reduces from $1437$ to $76$. 
%-------------------

Let us call \textit{Case I} the four-body basis used up to this stage, i.e., it contains all bound two-body correlated states, and \textit{Case II} the truncated four-body basis. This last one includes all correlated proton states and the correlated two-neutron states up to $10$ MeV of excitation energy. The strength of proton-neutron interaction was readjusted ($\chi_{CB}=7.9809$) to reproduce the experimental ground state energy of $^{44}$Ti. Table \ref{table.wfb_comparative} shows the energies and the amplitudes from the two four-body bases for the states $0^+_3$ and $0^+_5$. Both, the energies and amplitudes are very similar in the truncated basis. It is an open question how these differences translate to the calculation of the formation amplitude and the spectroscopic factor, this calculation is in progress. The following section analyzes the effect of truncation on the single particle width.
%----------------------
\begin{table}[h!t]
\centering
\caption{\label{table.wfb_comparative} Comparison of the $0^+_3$ and $0^+_5$ states from two different four-body bases. One includes all bound two-body correlated states, \textit{Case I} and the other with a reduction of the two-neutron bases states, \textit{Case II}.}
\begin{tabular}{c|cc|cc}
    \hline
    & \multicolumn{2}{c|}{$0^+_3$}  
    & \multicolumn{2}{c}{$0^+_5$}  \\ 
    Four-body basis 
    & \textit{Case I} & \textit{Case II} 
    & \textit{Case I} & \textit{Case II} \\ 
    \hline
    $E(MeV)$
    & $-26.782$ & $-26.742$
    & $-25.232$ & $-25.203$ \\[0.3mm]
    \hline
    $\mathcal{Z}^2_{0^+}$ 
    & $0.222$ & $0.221$ 
    & $0.240$ & $0.235$ \\
    $\mathcal{Z}^2_{2^+}$ 
    & $0.424$ & $0.431$ 
    & $0.157$ & $0.155$ \\
    $\mathcal{Z}^2_{4^+}$ 
    & $0.032$ & $0.032$ 
    & $0.245$ & $0.252$\\
    $\mathcal{Z}^2_{6^+}$ 
    & $0.296$ & $0.293$ 
    & $0.352$ & $0.352$ \\
    \hline
    $\sum_J \mathcal{Z}^2_J$
    & 0.974 & 0.977 
    & 0.994 & 0.994 \\
    \hline
\end{tabular}
\end{table}

%======================================
\subsection{Two-body decay width}\label{sec.t}
The alpha-decay half-lives calculation from the many-body spectroscopic factor requires the single-particle width and the formation amplitude \cite{2012IdBetan}. In this section, we calculated the single-particle widths for the $0^+_3$ and $0^+_5$ states. We use the well known current expression \cite{1961Humblet}
\begin{equation}\label{eq:penfact}
  \Gamma_{sp} = \frac{\hbar^2 \Re(k)}{\mu} 
     \frac{\left|u(r)\right|^2}
        {\left|H^+_0(\eta,kr)\right|^2} \, , 
\end{equation}
with $k$ the wave number related to the single-particle energy $\varepsilon$ defined below, $\mu$ the reduced mass of the $\alpha-^{40}$Ca system, $u(r)$ the corresponding relative Gamow wave function, and $H^+_0$ the outgoing Coulomb function. 

The single-particle wave function is calculated with the geometric parameters of Table \ref{table.vsp}. The strength of the Woods-Saxon is adjusted to reproduce the corrected energy of the $0^+$ states (Table \ref{table.wfb}, last column) relative to the threshold,
\begin{equation}\label{eq:tbalpha_energies}
    \varepsilon_i=E(0^+_i)-E(0^+_1) + Q_\alpha + \Delta E_{sc}
\end{equation}
with $Q_\alpha=-5.127$ MeV and $\Delta E_{sc}=4.7\si{keV}$ the electron screening. 

Table \ref{tab:spalpha_calcenergies} shows the resonant parameters (energy and width) for the alpha decay of the excited states $0^+_3$ and $0^+_5$ of $^{44}$Ti. The results of non-truncated (\textit{Case I}) and truncated bases (\textit{Case II}) are given.
%--------------------
\begin{table}[h!tb]
    \centering
    \caption{\label{tab:spalpha_calcenergies} Single-particle alpha decay energy, width (in MeV), and half-lives.}
    \begin{tabular}{c|ccc|ccc}
        \hline
        & \multicolumn{3}{c}{\textit{Case I}}
        & \multicolumn{3}{|c}{\textit{Case II}} \\
        \hline
        State & $\varepsilon$ & $\Gamma_{\rm{sp}}$ & $T_{sp}$ 
              & $\varepsilon$ & $\Gamma_{\rm{sp}}$ & $T_{sp}$ \\
        \hline
        $0^+_3$ & 1.519 & $0.581\sn{-12}$ & $0.786\si{ns}$ 
                & 1.559 & $1.261\sn{-12}$ & $0.362\si{ns}$ \\
        $0^+_5$ & 3.069 & $0.193\sn{-4}$ & $0.024\si{fs}$ 
                & 3.098 & $0.232\sn{-4}$ & $0.020\si{fs}$ \\
        \hline
    \end{tabular}
\end{table}

An alternative mean-field for the calculation of the single-particle width is provided by the local potential used in Ref. \cite{yamayaPhys.Rev.C1993} which describes the $\alpha$-elastic scattering from $^{40}$Ca: $a=1.04$ fm, $r_0=0.68$ fm. Using this potential the widths are reduced by a factor around two. For example, the widths (for the non-truncated basis) changed from $0.58\sn{-12}$ to $0.26\sn{-12}$ MeV, and from $0.19\sn{-4}$ to $0.11\sn{-4}$ MeV, for the $0^+_3$ and $0^+_5$ states, respectively. In any case, the $0^+_5$ state must be dismissed as a resonance; it is an example of the so-called wide resonance. In terms of the half-lives $T_{sp}=\hbar \ln 2/\Gamma_{sp}$, this resonance decays in $2.4\sn{-17}$ sec (or $4.2\sn{-17}$ sec with the mean-field of Ref. \cite{yamayaPhys.Rev.C1993}), which is orders of magnitude faster than the prompt gamma decay. 

The half-life $T_{sp}=0.786$ ns of the $0^+_3$ is a lower limit estimation (this figure changes to $1.746$ ns using the mean-field of Ref. \cite{yamayaPhys.Rev.C1993}) since the consideration of the alpha's structure, coded in the four-body spectroscopic factor $\mathcal{S}$, may increase this value ($T_{1/2}=T_{sp}/\mathcal{S}$)\cite{2012IdBetan}. Truncation of the four-body basis affects the single-particle half-lives by a factor of two. Even when this change may be significant, the reduction of a factor around twenty of the multidimensional integral is worth emphasizing.

%============================================
\section{Conclusions} \label{sec.conclusions}
The ground and $0^+$ excited states \textbf{of $\nn{44}{Ti}$} were studied using an effective interaction, and considering the correlation between all pairs of nucleons. 
The influence of the continuum spectrum was assessed using the  Gamow Shell Model framework. Our calculation supports the $0^+$ assignment for the experimental $(0,2)^+$ state at $6810(60)$ keV. The analysis of the collectivity of the excited states suggests that this state may be a candidate for alpha decay with a lower limit around the nanoseconds. \textbf{Although this value} is small to be measured, the structure of the two-neutron two-proton in the mother nucleus may increase this value through the spectroscopic factor. 

The calculation of the $^{44}$Ti alpha half-life using the four-body spectroscopic factor, including the full continuum and all pair of correlations, is undergoing.

% % % % % % % % % % % % 
\begin{acknowledgments}
Discussions with Witold Nazarewicz are gratefully acknowledged. We also thank Alex Brown and Nicolas Michel for their help benchmarking our Shell Model code. This work has been partially supported by the Consejo Nacional de Investigaciones Cient\'{\i}ficas y T\'ecnicas PIP-0930 (\hbox{CONICET}, Argentina). The computations were performed on the Computational Center of CCT-Rosario and CCAD-UNC, members of the SNCAD, MincyT-Argentina.
\end{acknowledgments}

%-------------
% Bibliography
%\bibliography{BibAlpha2020}

%-------------
\end{document}